\documentclass{article}

\usepackage{arxiv}
\renewcommand{\headeright}{}
\renewcommand{\undertitle}{}

\usepackage[T1]{fontenc}
\usepackage{mathptmx}

\usepackage{amsthm,amsmath,amsfonts,amssymb}
\usepackage[authoryear]{natbib}
\setcitestyle{authoryear,open={(},close={)},aysep={}}
\usepackage{graphicx}
\usepackage{longtable}
\usepackage{subcaption}
\usepackage[colorlinks,citecolor=blue,urlcolor=blue,linkcolor=black]{hyperref}
\usepackage{cleveref}

\graphicspath{{figures/}}

\title{The Persistent Non-Response Bias in a Sample-Matched Poll for the 2024 U.S. Presidential Election}

\author{%
  Jay Chooi\\
  Harvard University\\
  \texttt{jeqin\_chooi@college.harvard.edu}
}

\date{}

\begin{document}

\maketitle

\begin{abstract}
Donald Trump won the 2024 US Presidential Election despite polls predicting a Democratic lead, echoing the polling miss in 2016. Using the data defect correlation framework of \citet{mengStatisticalParadisesParadoxes2018}, we revisit the 60,000-respondent Cooperative Election Study and find that non-response bias for Trump voters persists on the same order of magnitude (\(\rho=-0.0030\) vs \(-0.0045\) in 2016) even under sample-matching to the US adult population. We additionally find evidence of positive response bias for Harris voters after adjusting for turnout. Consistent with findings in 2016, polling errors scale with state population size, and larger samples produce greater departures from conventional confidence intervals, with reductions of effective sample size exceeding 99\% in the largest states. We propose a pre-election bias correction estimator informed by historical data defect correlations and turnout rates that decreases RMSE from 0.13 to 0.05 using only prior election data, comparable to post-election weighting (RMSE 0.09).
\end{abstract}

\keywords{non-response bias \and data defect correlation \and election polling \and big data paradox \and sample matching}

\section{Introduction}
\label{sec1}

Polling has emerged as a highly sought-after method to forecast election outcomes, with 1,702 polls conducted for the 2024 US Presidential Election alone \citep{fivethirtyeightStatePolls20242024}. However, polls are estimators of the actual voting results and have errors. Some of these discrepancies between polls and actual elections can be attributed to the selection mechanism of the poll surveys, where the gold standard of simple random sampling (SRS) is often practically infeasible due to respondent refusals. Previous work has shown the existence of non-response bias in the 2016 US presidential election polls, which was particularly severe for Trump \citep{mengStatisticalParadisesParadoxes2018}. One explanation, termed the ``Shy Trump Supporter'' hypothesis, claims that respondents prefer not to admit that they will vote for Trump due to perceived social undersirability \citep{coppockDidShyTrump2017}. In the 2024 US presidential election where Trump ran again for president, pollsters predicted a slight edge for Democratic candidate Kamala Harris, with an aggregation of 25 polls predicting a 1.6\% Democratic lead \citep{270towin2024PresidentialElection2024}. Instead, Trump won all seven swing states with a 1.5\% lead in two-party popular vote share \citep{federalelectioncommission2024PresidentialElection2025}, in contrast to the polls, as in 2016.

By replicating the methodology from \cite{mengStatisticalParadisesParadoxes2018}, we found similar non-response bias for Trump in an established 2024 poll. The non-response bias occurs even if the samples are matched to the voter distribution. We also found evidence of positive response-bias for Harris when adjusted for turnout (Section \ref{subsec:data-defect-correlation}). As in the 2016 election, errors are magnified under large population sizes, with larger errors for larger states (Section \ref{subsec:llp}). Furthermore, bigger samples tend to escape conventional confidence intervals (Section \ref{subsec:method-big-data-paradox}). We computed the effective sample sizes of the polls (Section \ref{subsec:effective-sample-size}), and found that the sample size for the biggest states suffered a >99\% percentage reduction. To alleviate non-response bias, we propose a bias-corrected estimator in Section \ref{subsec:save_the_polls_methods} using historical data defect correlations. We evaluated it on the 2024 election, finding that it reduced RMSE by half and achieving comparable results to post-election weighting (Section \ref{subsec:save_the_polls_results}).

\section{Methods}

Following the setup of \cite{mengStatisticalParadisesParadoxes2018}, we conduct our analysis on the Common Content Dataset from the Cooperative Election Study (CES) \citep{schaffnerCooperativeElectionStudy2025}. The CES data includes a nationally representative sample of 60,000 American adults interviewed before and after the 2024 elections, in September to October and November to December, respectively. The CES study employs two strategies to achieve representativeness: sample-matching and weighting.

\subsection*{Sample-matching.} CES sources its respondents from YouGov. The target distribution is all US adults, informed by the 2023 American Community Survey (ACS) \citep{acs2023_1yr}. For each sample drawn from this target distribution, i.e. a US adult, a sample from the YouGov pool of respondents is selected such that the sample is as identical as possible to that drawn from the target distribution. To measure identicalness, CES uses a weighted sum of Euclidean distances between individual dimensions like gender and race.

\subsection*{Weighting} As the samples drawn from ACS and YouGov might not be a perfect match, CES applies weighting to further improve representativeness. First, entropy balancing is used to weight samples to match moment conditions with the ACS (e.g. on gender and race and their interaction effects). Next, post-stratification is used to weight on more dimensions, including on the 2024 presidential vote choice. For validated voters, the weighting is computed again to be representative of the voter population. Large weights are trimmed.

Since the weights are calculated using the 2024 election results, applying weighting to the matched-samples will mechanically correct for any non-response bias \emph{after} the election. As we are interested in the polling results before the election, we investigate the extent of non-response bias in the unweighted matched-samples. The bias-corrected estimator we propose is also to be used before the election.

\subsection{Computing the data defect correlation}\label{subsec:data-defect-correlation}

The actual error between the sample mean and the population mean can be decomposed into three factors through the Meng equation\footnote{See \citet{baileyNewParadigmPolling2023} for an exposition on how the Meng equation could redefine the polling paradigm.} \citep[Equation 2.3]{mengStatisticalParadisesParadoxes2018}:

  \begin{equation}\label{eq:meng_identity}
\underbrace{\bar G_n - \bar G_N}_{\text{Sample Error}}
= \underbrace{\rho_{R,G}}_{\text{Data Quality}}
\;\times\;
\underbrace{\sqrt{\frac{1 - f}{f}}}_{\text{Data Quantity}}
\;\times\;
\underbrace{\sigma_G}_{\text{Problem Difficulty}}
  \end{equation}

  where
  \begin{align*}
    &\bar G_n\text{ is the sample mean, e.g. the share of respondents who prefers Trump,}\\
    &\bar G_N\text{ is the population mean, e.g. the vote share of Trump,}\\
    &\rho_{R,G}\text{ is the \emph{data defect correlation}, e.g. the correlation between responding to }\\
    &\text{the survey and voting for Trump,}\\
    &f\text{ is the ratio of sample size to population size,}\\
    &\sigma_G\text{ is the inherent standard deviation of }G.
  \end{align*}

Before the election, we only have estimate \(\bar G_n\). After the election, we will have observed \(\bar G_N, f,\sigma_G\). We can then proceed to calculate the data defect correlation.

\begin{equation}\label{eq:computing_data_defect_correlation}
\rho_{R,G}
=
\frac{\overline{G}_n - \bar{G}_N}
     {\sigma_{G}\,\sqrt{\dfrac{1 - f}{f}}}
\end{equation}

In the results (Section \ref{subsec:data-defect-correlation-results}), we plot a histogram of the data defect correlation across all states for both Trump and Harris. We assume that the data defect correlation is i.i.d. across states, such that a scenario with negligible non-response bias will have a distribution around 0, while a scenario with constant non-zero non-response bias will have a distribution centered away from zero. We found that Trump has a positive raw mean much larger than zero, while the raw mean for Harris is very close to zero.

\subsection{Testing for the Law of Large Populations}\label{subsec:llp}

One way to quantify the error under non-response bias is to report it in units of error under an otherwise perfect SRS. To do this, note that \(\bar G_n\) under SRS is unbiased, so its mean-squared error (MSE) is the same as its variance:

\begin{equation}\label{eq:srs_error}
\mathbb V_{\text{SRS}}(\bar{G}_n) = \frac{1 - f}{n} \frac{N}{N - 1} \sigma_G^2,
\end{equation}

Combining the SRS error in \Cref{eq:srs_error} with Meng identity in \Cref{eq:meng_identity} , we obtain a measure of the non-response error in units of SRS error:
\begin{equation}\label{eq:compute_znn}
    Z_{n,N}
\;\equiv\;
\frac{\bar{G}_n - \bar{G}_N}
     {\sqrt{\mathbb  V_{\mathrm{SRS}}(\bar{G}_n)}}
\;=
\sqrt{N - 1}\;\rho_{R,G}.
\end{equation}

The dependence of \(Z_{n,N}\) on \(\sqrt{N}\) under non-response bias is termed in \cite{mengStatisticalParadisesParadoxes2018} as the Law of Large Populations:

\begin{quote}
\textbf{Law of Large Populations (LLP)} Among studies sharing the same (fixed) average data defect correlation $\mathbb{E}_R(\rho_{R,G}) \neq 0$, the (stochastic) error of $\bar{G}_n$, relative to its benchmark under SRS, grows with the population size $N$ at the rate of $\sqrt{N}$.
\end{quote}

To test for this relationship, we have to  find evidence of a square root dependence on \(N\). Taking logs on both sides of \Cref{eq:compute_znn} yields
\begin{equation}\label{eq:znn_regression}
    \log\lvert Z_{n,N}\rvert
\approx\log\lvert {\rho}_{R,G}\rvert
+ 0.5\,\log N.
\end{equation}

Using the post-election error to compute \(Z_{n,N}\) through the left-hand side of \Cref{eq:compute_znn} and plotting it against \(\log N\), we expect to see a gradient of \(0.5\) if \(\rho_{R,G}\) stays roughly constant. In \Cref{subsec:llp-results}, we show that the errors for Trump indeed have a gradient of 0.5 when using all respondents or estimated likely voters.

\subsection{Lower Confidence in Bigger Data with the Big Data Paradox}\label{subsec:method-big-data-paradox}

We construct confidence intervals for \(\bar G_n\) by assuming a Normal distribution for the error, and finding \(Z_n\), the z-score. Note that the \(Z_n\) here is distinct from \(Z_{n,N}\) in Section \ref{subsec:llp}.

\begin{equation}
Z_n
= \frac{\bar{G}_n - \bar{G}_N}{\sqrt{\bar{G}_n(1 - \bar{G}_n)/n}}
= \frac{\sqrt{n}\,\sqrt{D_O}\,\rho_{R,G}}
       {\sqrt{1 - D_O\,\rho_{R,G}^2}
        - \sqrt{D_O}\,\rho_{R,G}
          \left(\sqrt{\frac{\bar{G}_N}{1 - \bar{G}_N}} - \sqrt{\frac{1 - \bar{G}_N}{\bar{G}_N}}\right)}.
\end{equation}
where \(D_O=\frac{1-f}{f}\). For tight races (i.e. \(\bar{G}_N\approx 0.5\)), the second term in the denominator is close to zero, yielding
\begin{equation}\label{eq:zn_tight_race}
    Z_n
\approx\sqrt{n} \cdot\frac{\sqrt{D_O}\rho_{R,G}}
       {\sqrt{1 - D_O\rho_{R,G}^2}}
\end{equation}

In this case, if we further hold the sample ratio \(f\) constant, and thus \(D_O\) constant, and when \(\rho_{R,G}^2\) is non-zero and does not depend on \(n\), then \(Z_n\) scales by \(\sqrt{n}\). This \(n^{1/2}\) relationship is in stark comparison to the usual confidence interval under SRS that scales instead with \(n^{-1/2}\). Consequently, when the selection mechanism differs from SRS and there is a correlation between selection and the type of response, we predict that samples with large \(n\) will fall outside typical confidence intervals. This is known in \cite{mengStatisticalParadisesParadoxes2018} as the \emph{Big Data Paradox}.

\begin{quote}
\textbf{The Big Data Paradox}: The bigger the data, the surer we fool ourselves.
\end{quote}

In \Cref{subsec:lower-confidence-in-bigger-data-with-the-big-data-paradox}, we indeed found that states with bigger samples have bigger errors when used to predict Trump's vote share.

\subsection{Computing the effective sample sizes}\label{subsec:effective-sample-size}

Another way to quantify the error is to calculate the \emph{effective sample size}, which is the sample size under SRS that will result in the same observed error in expectation. To solve for the effective sample size, we equate the mean-squared error (MSE) of \(\bar {G_n}\) to the error under SRS (which is the variance) using an effective sample size \(n_\mathrm{eff}\):
\begin{equation}
    \mathbb E_R[\bar G_n-\bar G_N]^2=\mathbb V_{SRS}(\bar G_{n_\mathrm{eff}})
\end{equation}

Using Jensen's inequality and the assumption that \(\mathbb E_R[\rho_{R,G}^2]\leq\frac{f}{1-f}\), \cite{mengStatisticalParadisesParadoxes2018} showed that
\begin{equation}\label{eq:effective-sample-size}
    n_{\mathrm{eff}}
\le \frac{1}{\mathbb E_R[\rho_{R,G}^2]}\frac{f}{1-f}
\end{equation}

Using the range of sample ratios (0.0002 to 0.0006) in CES, the assumption becomes \(\mathbb E_R[\rho_{R,G}^2]\leq \frac{0.0002}{1-0.0002}\approx 0.0002\). In \Cref{fig:fig-5-data-defect-corr-distribution}, we show that all data defect correlations computed satisfy \(\rho_{R,G}^2\leq 0.0001\), so the assumption holds comfortably. In Section \ref{subsec:effective-sample-size-results}, we compute the upper bound for the effective sample sizes of each state and compute the percentage reductions in sample sizes, which achieve \(>99\%\) in the biggest states.

\subsection{Saving the Polls with Prior-Informed Bias Correction}\label{subsec:save_the_polls_methods}

We propose an estimator of vote share by applying bias correction to the the sample average. Using the Meng identity (\Cref{eq:meng_identity}), we can theoretically construct an unbiased estimator
\begin{equation}
\bar G_n'=\bar G_n - \rho_{R,G}\times \sqrt{\frac{1-f}{f}}\times \hat\sigma_{G}
\end{equation}

This is ideal, but we do not have access to \(\rho_{R,G},f,\sigma_G\) before the election. We estimate each of the unknown quantities in turn. First, we estimate \(\rho_{R,G}\) using the average data defect correlation from a previous survey. In our results, we demonstrate this approach using the CES survery from 2016, \(\bar\rho_{R,G,2016}\). Then, we estimate \(f\) by using the turnout of each state from a previous election. To be consistent, we again use turnout data from 2016 but applying that to the voting-eligible population in 2024, \(N_\mathrm{VEP,2024}\).
\begin{equation}
    \hat f_{2024}=\frac{n}{N_\mathrm{VEP,2024}\cdot t_{2016}}
\end{equation}

Finally, we estimate the problem difficulty \(\sigma_G\) by taking the sample standard deviation \(\hat\sigma_{G,2024}\) from the current polls being analyzed. This yields our estimator in Equation \ref{eq:bc-estimator}:

\begin{equation}\label{eq:bc-estimator}
\bar G_n^*=\bar G_n - \bar \rho_{R,G,2016}\times \sqrt{\frac{1-\hat f_{2024}}{\hat f_{2024}}}\times \hat\sigma_{G,2024}
\end{equation}

Note that this estimator can be computed pre-election, and more sophisticated estimators for each of the components \(\rho,f,\sigma\) could be possible. We evaluated the performance of this estimator in Section \ref{subsec:save_the_polls_results} by using the datasets from the Election Lab at the University of Florida for turnout in 2016 \citep{mcdonald2016GeneralElection2023} and for the voting-eligible population in 2024 \citep{mcdonald2024GeneralElection2024}.

\section{Results}\label{sec:results}

\begin{figure}[!htb]
\centering
\includegraphics[width=\linewidth]{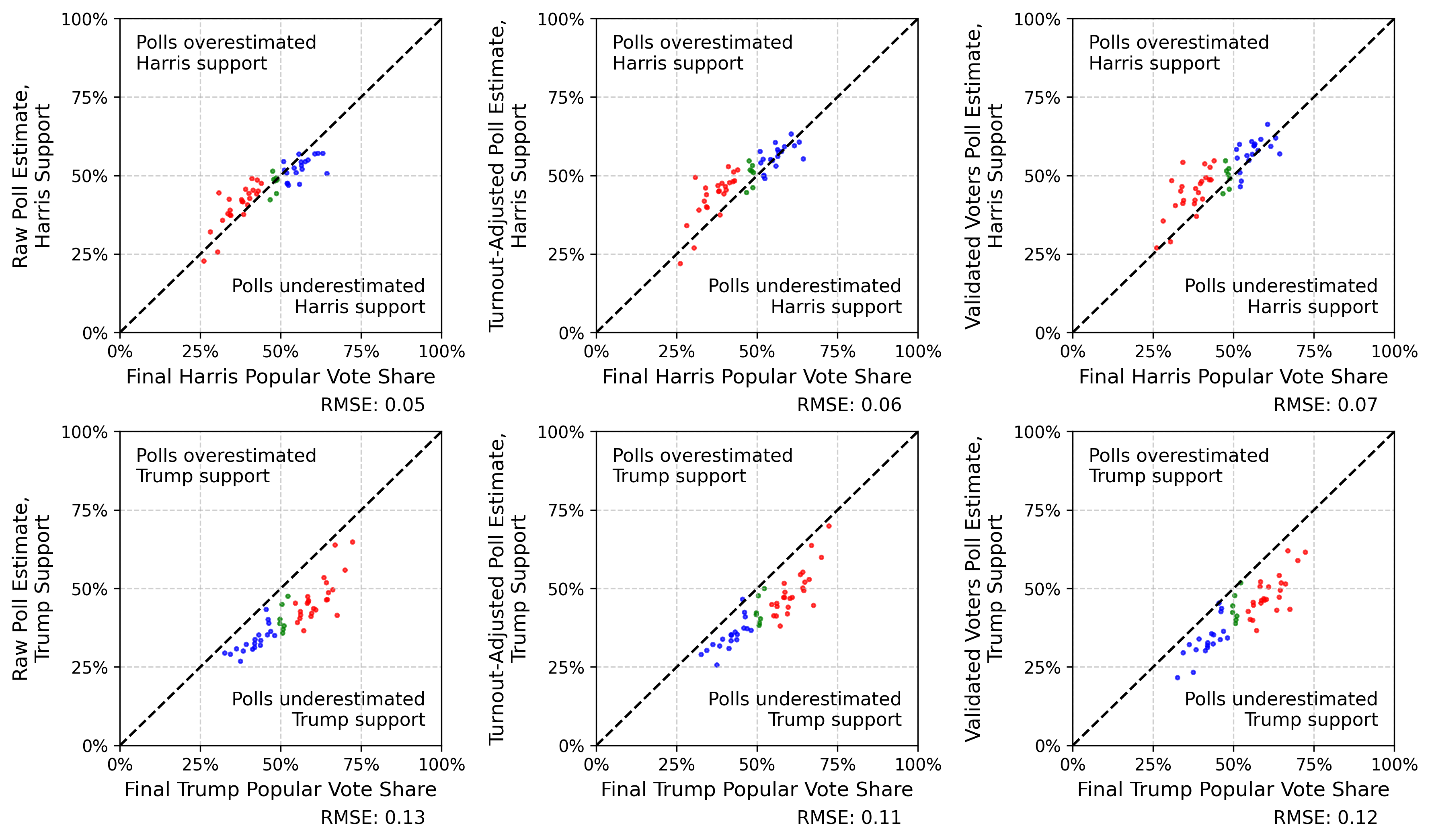}
\caption{Poll estimate vs actual vote share for Harris and Trump for each state, using raw polls and turnout-adjusted polls. Red states are solid or likely Republican states, blue states are solid or likely Democrat states, green states are tossup states. Classifications are from the 2024 CPR Electoral College Ratings \citep{cookpoliticalreport2024CPRPresident2024}. Details on turnout estimation can be found in Appendix \ref{subsec:turnout-estimate-details}.}
\label{fig:fig-4-poll-estimate}
\end{figure}

In Figure \ref{fig:fig-4-poll-estimate}, we see that the poll estimates for Harris are roughly in line with the actual vote share, with a RMSE of 0.05, 0.06, 0.07 for raw polls, turnout-adjusted polls and validated voter polls respectively. For Trump, there is visible underestimation even after accounting for voter validation, with this pattern holding across all states. The RMSE is higher at 0.13, 0.11 and 0.12. This points towards the existence of non-response bias from Trump voters that is more serious than non-response bias from Harris voters. In other words, the survey when using only sample-matching under-represents voters who vote for Trump.

When post-election CES weighting is applied, the RMSE for Trump decreases down to Harris' level across all sample populations, removing non-response bias (\Cref{fig:figure_4_weighted}).

\begin{figure}[!htb]
    \centering
       \begin{subfigure}{\linewidth}
\caption{Using all respondents}
\includegraphics[width=\linewidth]{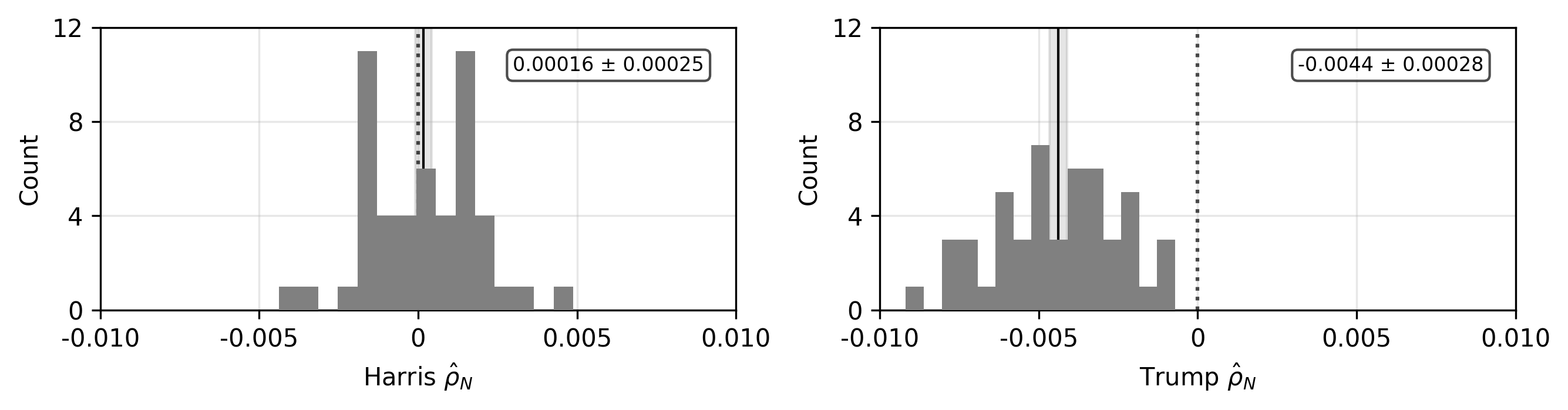}
\end{subfigure}

    \begin{subfigure}{\linewidth}
\caption{Using likely voters (pre-election)}
\includegraphics[width=\linewidth]{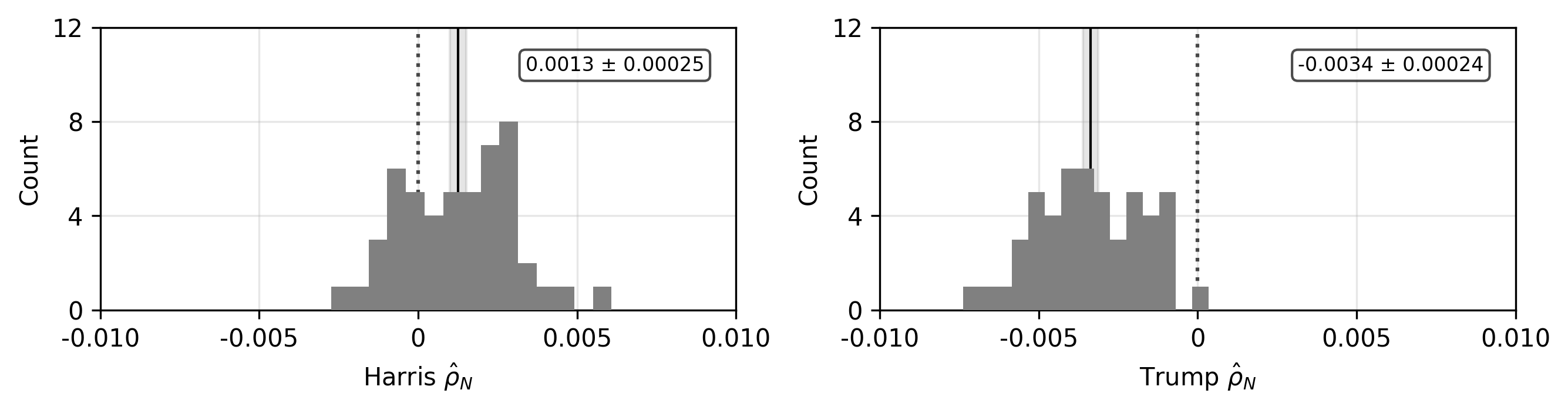}
\end{subfigure}

       \begin{subfigure}{\linewidth}
\caption{Using validated voters (post-election)}
\includegraphics[width=\linewidth]{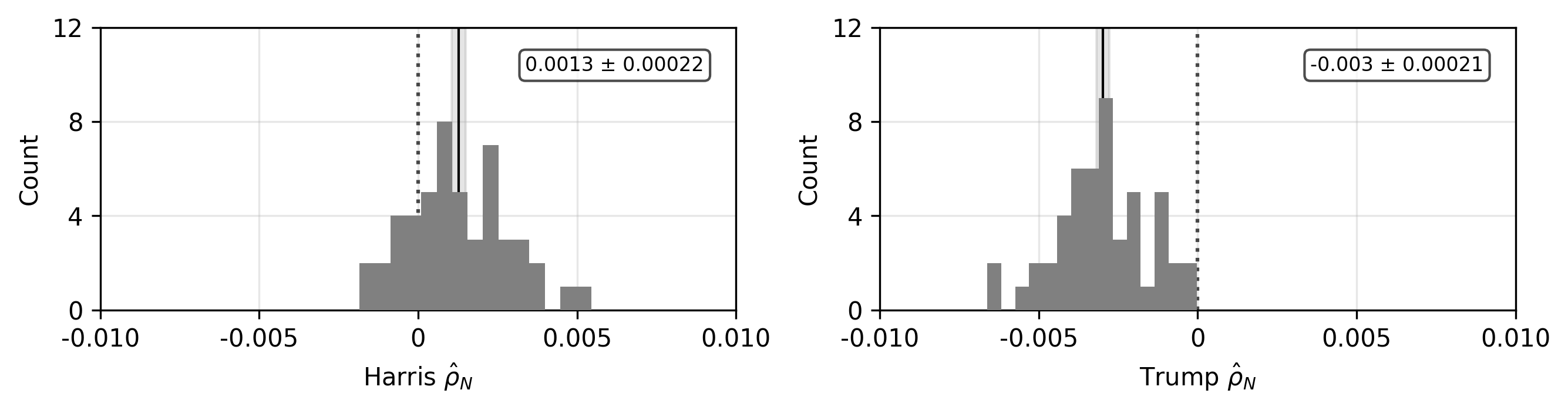}
\end{subfigure}

    \caption{Histogram of data defect correlation for Harris and Trump voters for each state in the 2024 US presidential election. The boxes in the top right of the plots show ``mean \(\pm 2\) standard error''. Zero data defect correlation is marked in dotted vertical lines. The means are marked in solid vertical lines. \(\pm 2\) standard errors are shaded in light gray.}
    \label{fig:fig-5-data-defect-corr-distribution}
\end{figure}

\subsection{Computing the data defect correlation}\label{subsec:data-defect-correlation-results}

\Cref{fig:fig-5-data-defect-corr-distribution} plots the data defect correlation for each state using \Cref{eq:computing_data_defect_correlation} with three subsets of data: all respondents, likely voters, and validated voters. Likely voters are same as those used for turnout-adjusted estimates in Figure \ref{fig:fig-4-poll-estimate}. Using all respondents, there appears to be no data defect correlation for Harris, while Trump has a data defect correlation centered around \(-0.0044\). However, after correcting for turnout using either estimates pre-election or validated voters post-election, the data defect correlations for Harris moved away from non-zero to a positive 0.0013 and for Trump the data defect correlations decreased to \(-0.0030\) (\cite{mengStatisticalParadisesParadoxes2018} reports \(-0.0045\)). In all three cases, CES exhibits non-response bias for Trump voters. In turnout-adjusted cases, CES exhibits positive response bias for Harris. The combination of these two biases both contribute to an overestimation of Harris' support.

\subsection{Testing for the Law of Large Populations}\label{subsec:llp-results}

\begin{figure}[!htb]
    \centering

\begin{subfigure}{\linewidth}
\caption{Using all respondents}
\includegraphics[width=\linewidth]{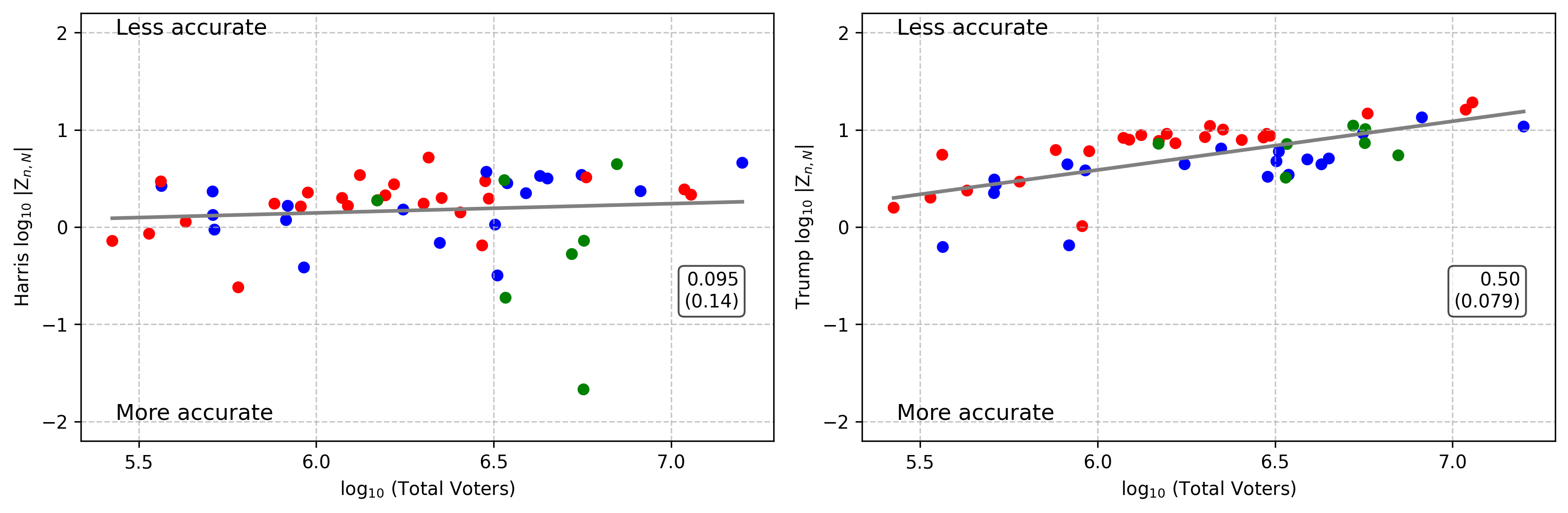}
\end{subfigure}

    \begin{subfigure}{\linewidth}
\caption{Using likely voters (pre-election)}
\includegraphics[width=\linewidth]{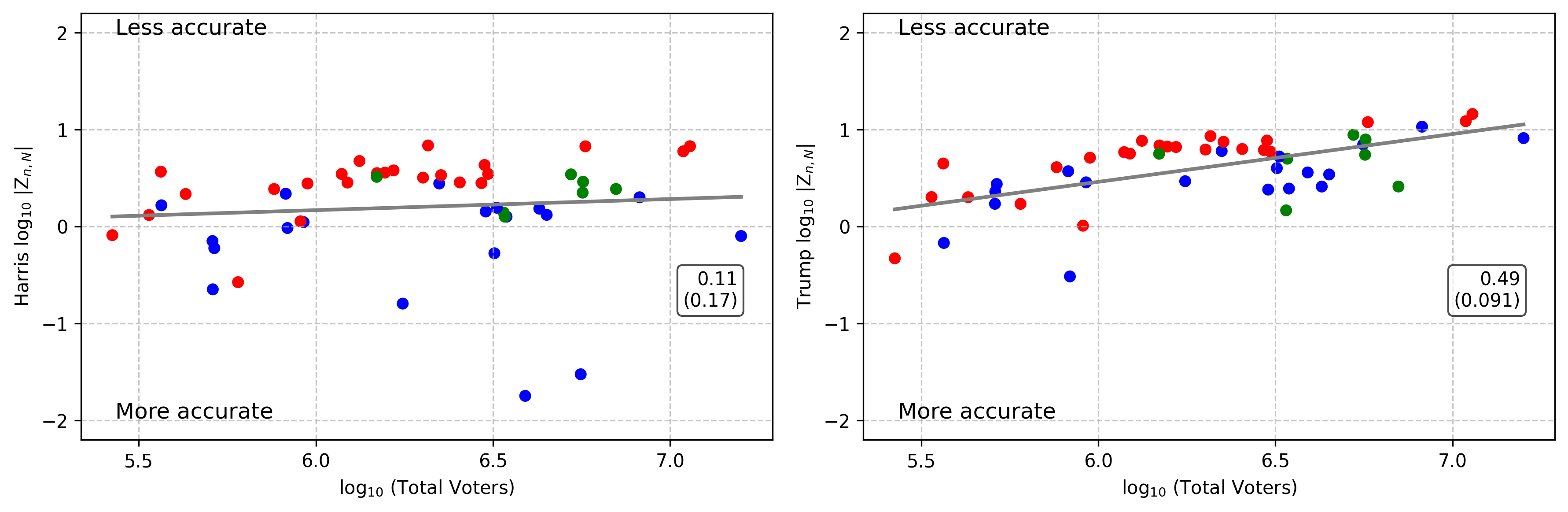}
\end{subfigure}

       \begin{subfigure}{\linewidth}
\caption{Using validated voters (post-election)}
\includegraphics[width=\linewidth]{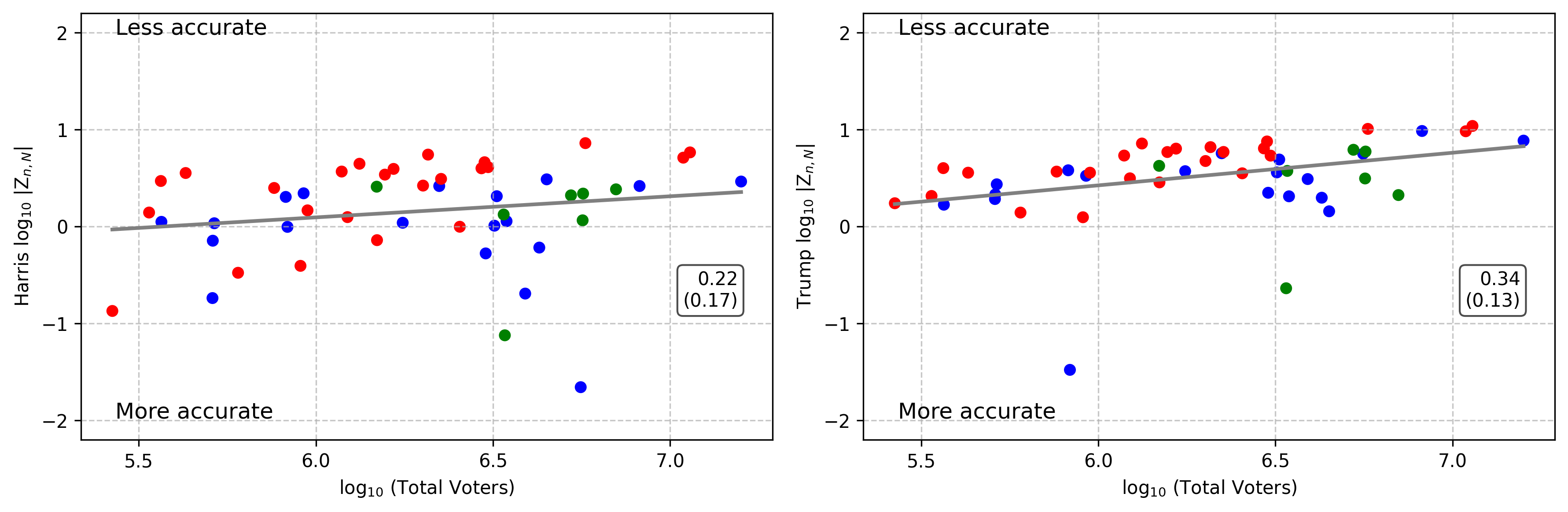}
\end{subfigure}

\caption{Error in SRS-units \(Z_{n,N}\) to the number of total votes in each state.}
\label{fig:fig-6-llp}
\end{figure}

In Figure \ref{fig:fig-6-llp}, we plot \(|Z_{n,N}|\), the relative error \(\bar G_n-\bar G_N\) in units of the SRS error \(\mathbb V_{SRS}(\bar G_n)\), to the number of total votes recorded in each state (including votes for each candidate and other alternative candidates), with log axes. We also fitted the regression of \Cref{eq:znn_regression} to the plot. We report the gradient coefficient of \(\log N\) and its standard error in the boxes.

We analyze the case for all respondents and for likely voters. For Harris, we found that \(|Z_{n,N}|\) stays roughly constant regardless of the number of voters in each state, which echoes the findings for Democrats in 2016 in \cite{mengStatisticalParadisesParadoxes2018}. The standard error for the gradient for Harris puts zero within range, which is expected for a SRS where \(Z_{n,N}\) stays relatively constant over \(N\). For Trump, we see a visible increase in \(|Z_{n,N}|\) as the actualized voter population size grows. We found the gradient coefficient considered alongside the standard error to be well within 0.5 and away from zero. Recalling the construction of \Cref{eq:znn_regression}, the clear linear trend in the right plot of Fig \ref{fig:fig-6-llp} with a gradient of 0.5 is indicative of a non-zero data defect correlation that does not depend on \(N\), while the estimators \(\bar G_n\) for Trump's vote share have errors that scale with the state population size \(\sqrt{N}\). In Figure 6 of \cite{mengStatisticalParadisesParadoxes2018}, a gradient close to 0.5 (0.44 with a standard error of 0.09) was also reported for Trump. For post-election validated voters, the gradient for Harris is no longer close to zero within standard errors, and the gradient the Trump is no longer closer to 0.5 within standard errors.

\subsection{Lower Confidence in Bigger Data with the Big Data Paradox}\label{subsec:lower-confidence-in-bigger-data-with-the-big-data-paradox}

In Section \ref{subsec:method-big-data-paradox}, we assumed constant sample ratios to describe a dependence of the standardized error \(Z_n\) on \(\sqrt{n}\). Although the sample ratios in CES are not constant across states, they do not vary as widely as much as the sample sizes for each state (\Cref{fig:sample-size-ratio-scatterplot}; \Cref{tab:sample-details}). Approximating the sample ratios to be constant across states, we then check for the \(n^{1/2}\) dependence described in Section \ref{subsec:method-big-data-paradox}.

\begin{figure}[!htb]
\centering
\includegraphics[width=\linewidth]{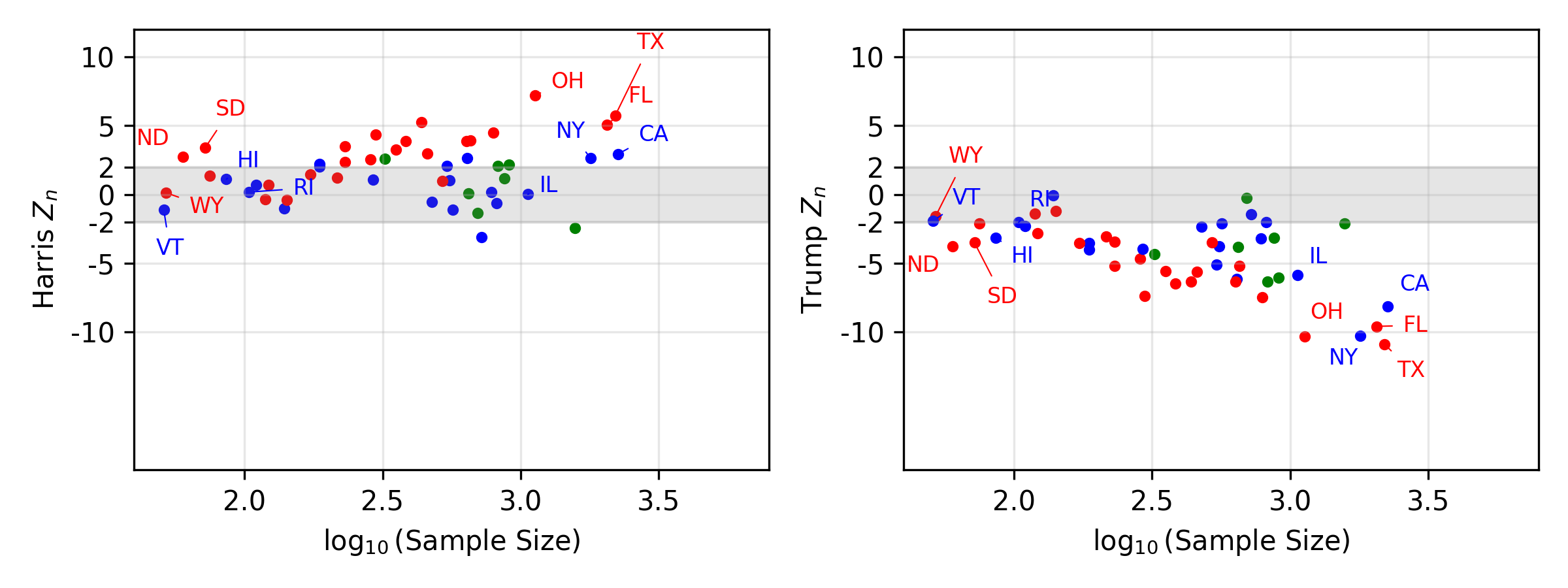}
\caption{Standardized error \(Z_n\) to sample size for each state, using validated voters.}
\label{fig:fig-7-big-data-paradox}

\end{figure}

In Figure \ref{fig:fig-7-big-data-paradox}, we plot \(Z_n\), the error standardized by its error, to the log of the sample size \(\log n\). We shade the confidence intervals of \(\pm 2\) standard deviation under typical Normal distribution assumptions in gray. We found that the errors for Trump tends to exit the confidence interval as the sample size grows, reaching more than 10 standard deviations for states with the biggest samples (e.g. California and Texas). The states with smaller samples (e.g. Vermont and Wyoming) are within the confidence intervals. For Harris, we also found that some states with bigger samples exit the confidence intervals, though the transgression is not as severe and numerous as compared to Trump.

\subsection{Computing the effective sample sizes}\label{subsec:effective-sample-size-results}

Using the methods in Section \ref{subsec:effective-sample-size}, we compute the effective sample size for each state, which is the sample size that yields the same error under SRS as the errors observed with the current selection mechanism in CES. We include all respondents for this section, not just validated voters. We report the results for the five biggest states by actualized voters in Table \ref{tab:big-eff-sample-size} and for the five smallest states in Table \ref{tab:small-eff-sample-size}. We report the full results for all states in \Cref{tab:effective-sample-sizes}.

\begin{table}[!tb]
\caption{Effective sample sizes and percentage reduction for predicting Trump vote shares in CES 2024 for the five biggest states}
\label{tab:big-eff-sample-size}
\centering
\begin{tabular}{lrrrr}
\hline
 State        &   Sample size & Total votes   &   Effective sample size& Percentage reduction   \\
\hline
 Texas        &         4222 &   11,380,105 &                             19 &                        99.55\% \\
 California   &         4035 &   15,862,536 &                             13 &                        99.67\% \\
 Florida      &         3543 &   10,893,548 &                             17 &                        99.53\% \\
 New York     &         3205 &    8,199,062 &                             20 &                        99.37\% \\
 Pennsylvania &         2503 &    7,034,206 &                             18 &                        99.27\% \\
\hline
\end{tabular}
\end{table}

\begin{table}[!tb]
\caption{Effective sample sizes and percentage reduction for predicting Trump vote shares in CES 2024 for the five smallest states}
\label{tab:small-eff-sample-size}
\centering
\begin{tabular}{lrrrr}
\hline
 State        &   Sample size & Total votes   &   Effective sample size& Percentage reduction   \\
\hline
 Wyoming      &            88 & 266,353       &                              17 & 80.60\%                         \\
 Alaska       &           117 & 338,177       &                              18 & 84.72\%                         \\
 North Dakota &            99 & 365,059       &                              14 & 85.85\%                         \\
 Vermont      &            85 & 366,389       &                              12 & 85.90\%                         \\
 South Dakota &           131 & 428,922       &                              16 & 87.95\%                         \\
\hline
\end{tabular}
\end{table}

We found that the effective sample sizes are between 12 and 22 (inclusive), even though the actual sample sizes are between 85 and 4,222. The biggest percentage reductions occur for the biggest states, and the five biggest states all have percentage reductions above 99\%. Meanwhile, the five smallest states have percentage reductions below 90\%, and are also the five states with the smallest percentage reductions. Examining \Cref{eq:effective-sample-size}, it is expected that effective sample size stays constant regardless of actual sample size when the sample ratio stays constant (which is roughly the case here; see \Cref{fig:sample-size-ratio-scatterplot} and \Cref{tab:sample-details}). Since the effective sample size is roughly constant across states in the CES dataset, it follows that the percentage reduction is the biggest in the biggest states.

\subsection{Saving the Polls with Bias Correction}\label{subsec:save_the_polls_results}

\begin{figure}[!htb]
\centering
\includegraphics[width=.5\linewidth]{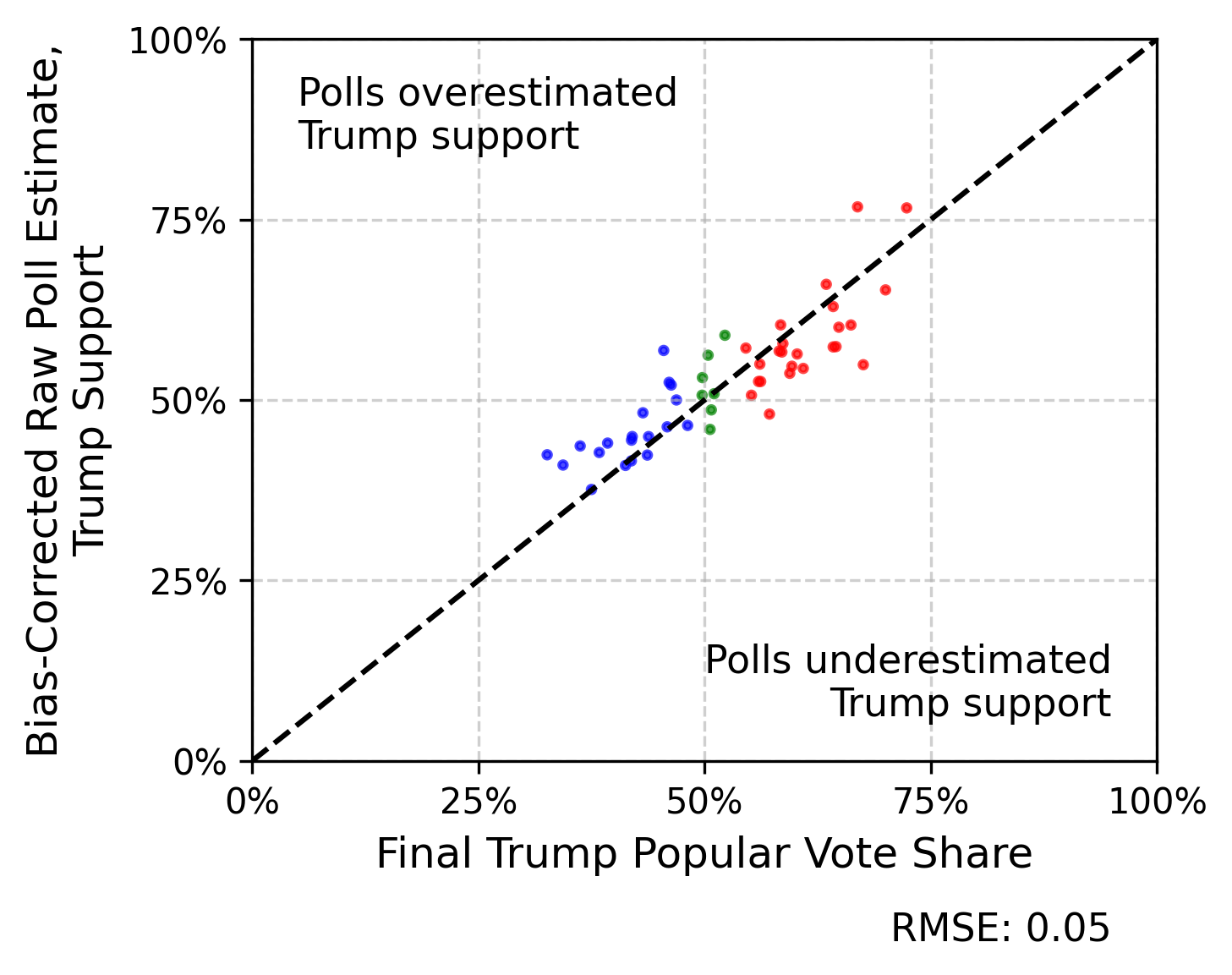}
\caption{Bias-corrected raw poll estimate of Trump's vote share in 2024. There is no consistent underestimation of Trump's performance, and the RMSE is halved from that of the raw matched-samples (Figure \ref{fig:fig-4-poll-estimate}) and similar to when post-election weights are applied (\Cref{fig:figure_4_weighted}).}
\label{fig:bias-correction}

\end{figure}

We computed the bias-corrected estimates for 2024 using \Cref{eq:bc-estimator} and plotted it against the actualized vote share in Figure \ref{fig:bias-correction}. There is no longer any visible and consistent underestimation of Trump's vote share across states compared to \Cref{fig:fig-4-poll-estimate}. Furthermore, the RMSE is improved down to 0.05, which is less than half of what was observed for Trump (0.13) if we use the sample mean. Note that the good performance of the new estimator \(\bar G_n^*\) is likely to be driven by the similarity between the average data defect correlations for 2016 and 2024 (for validated voters, it is \(-0.0030\) vs \(0.0045\), though we used all respondents here). For future elections, the average data defect correlation might shift, especially if CES makes changes to their sampling strategy, and thus might vary in utility.

\section{Discussion}\label{sec:discussion}

Before the election, using either the all respondents subset or the likely voters subset revealed a strong non-response bias from Trump voters, with the effect diminishing slightly when analyzing post-election validated voters, and mostly eliminated after applying post-election weights (\Cref{fig:figure_4_weighted}). This raises questions on how we can improve polling techniques so that non-response bias could be alleviated in pre-election analysis. Most often before the elections, pollsters will apply weights to correct for non-response bias. However, \citet{cohnWeGaveFour2016} showed that different reasonable weighting choices on the same raw data result in highly different forecasts. \citet{dominitzUsingTotalMargin2024} proposes accounting for uncertainty on the leanings of non-respondents, which leads to an error measure of \textasciitilde49\% on a poll that had a reported error of \textasciitilde3\%. Furthermore, \citet{baileyNewParadigmPolling2023} argued that the ignorability assumption for weighting could often times be violated.

To reduce the data defect correlation, we outlined a method using corrections informed by previous surveys and elections in Section \ref{subsec:save_the_polls_methods}, and there is rich literature on more sophisticated correction mechanisms or alternative methods. \citet{Bailey_2024} demonstrated a method to calculate data defect correlations using randomized response instruments on an Ipsos poll. For the 2020 election, \citet{isakovPrincipledUnskewingViewing2020} approached this problem similarly using the Meng equation and computed bias-corrected estimates before the election. To adjust for non-response, \citet{cohn2023polling} sampled much more from groups predicted to have high non-response rates, even though this comes as a significant polling expense. The problem of non-response could also be approached from a Bayesian angle \citep{gelmanGrapplingUncertaintyForecasting2024}.

We assume that survey respondents answer truthfully. We also allow the assumption that a negligible number of voters could have changed their minds and voted differently than the candidate they responded with in the survey. We note that there are respondents who did not express their preferences in the survey. Out of \(50{,}396\) survey respondents who were asked the question ``Which candidate for President of the United States do you prefer?'', \(2{,}999\) respondents responded ``I'm not sure''. Even if we assign \emph{all} these responses to count towards Trump's estimate, we still get a consistent underestimation of Trump's share across all states. We report this result in \Cref{fig:not-sure-is-trump}. This shows that non-response bias still remains in the sample-matching mechanism of CES even after accounting for dishonest responses.

One idea to counteract a high data defect correlation is to increase the overall data quantity, as hinted at in the Meng identity (\Cref{eq:meng_identity}). However, as argued in Section 3 in \cite{mengStatisticalParadisesParadoxes2018}, this is not feasible. We present a new example using the 2024 election. Suppose we would like an effective sample size of 1,000 in Pennsylvania, a swing state which had 7 million votes counted in 2024 and where CES collected 2,503 survey responses. Using \Cref{eq:effective-sample-size}, which gives a theoretical upper bound to the effective sample size, we require a sample ratio of at least \(0.019\), which is approximately \(130{,}000\) Pennsylvanian respondents. The entire CES survey only has \(60,000\) respondents across all states. To move forward and improve polling estimates, data quantity alone could not be relied on to fully overcome the issues from data quality: non-response bias itself has to be directly addressed.

To analyze the extent of non-response bias in other surveys, one future work would be to compute the data defect correlation for other nation-wide state-level surveys, including that of \cite{270towin2024PresidentialElection2024}, and compare it to CES.

\section*{Acknowledgments}
The author thanks Xiao-Li Meng, Kyla Chasalow, Shiro Kuriwaki, and Michael Isakov for their many helpful comments. This work received no funding.

\bibliographystyle{plainnat}
\bibliography{references}

\clearpage
\appendix
\counterwithin{figure}{section}
\counterwithin{table}{section}
\begin{center}
{\LARGE\bfseries Appendix}
\end{center}
\vspace{1em}

\section{Data Availability Statement}
\label{sec:data-availability}

All data used in this study are publicly available from third-party sources:

\begin{itemize}
\item \textbf{Cooperative Election Study (CES) 2024 Common Content.} The 60{,}000-respondent survey analyzed throughout this paper (file \texttt{CCES24\_Common\_OUTPUT\_vv\_topost\_final.csv}) is distributed by CES on the Harvard Dataverse at \url{https://dataverse.harvard.edu/dataset.xhtml?persistentId=doi:10.7910/DVN/X11EP6} \citep{schaffnerCooperativeElectionStudy2025}. Users should register with Harvard Dataverse and download the Common Content file directly.

\item \textbf{2024 state-level election results.} Certified state-level vote totals are from the Federal Election Commission \citep{federalelectioncommissionfecElectionResultsVoting2024}.

\item \textbf{Turnout and voting-eligible population.} State-level turnout rates and voting-eligible population estimates for 2016 and 2024 are from the Election Lab at the University of Florida at \url{https://election.lab.ufl.edu/dataset/} \citep{mcdonald2016GeneralElection2023, mcdonald2024GeneralElection2024}.

\item \textbf{Pre-election state classifications.} The solid/likely/tossup classifications used to color states in \Cref{fig:fig-4-poll-estimate} are from the 2024 Cook Political Report Electoral College Ratings \citep{cookpoliticalreport2024CPRPresident2024}.
\end{itemize}

All intermediate files are produced by the analysis code from the raw sources above. The analysis code that reproduces every figure, table, and number in this paper is openly available at \url{https://github.com/jeqcho/data-defect-2024-us-election}.

\section{Applying Post-Election Weights}
\label{sec:using-weighted-samples}

\begin{figure}[!htb]
  \centering
  \includegraphics[width=\linewidth]{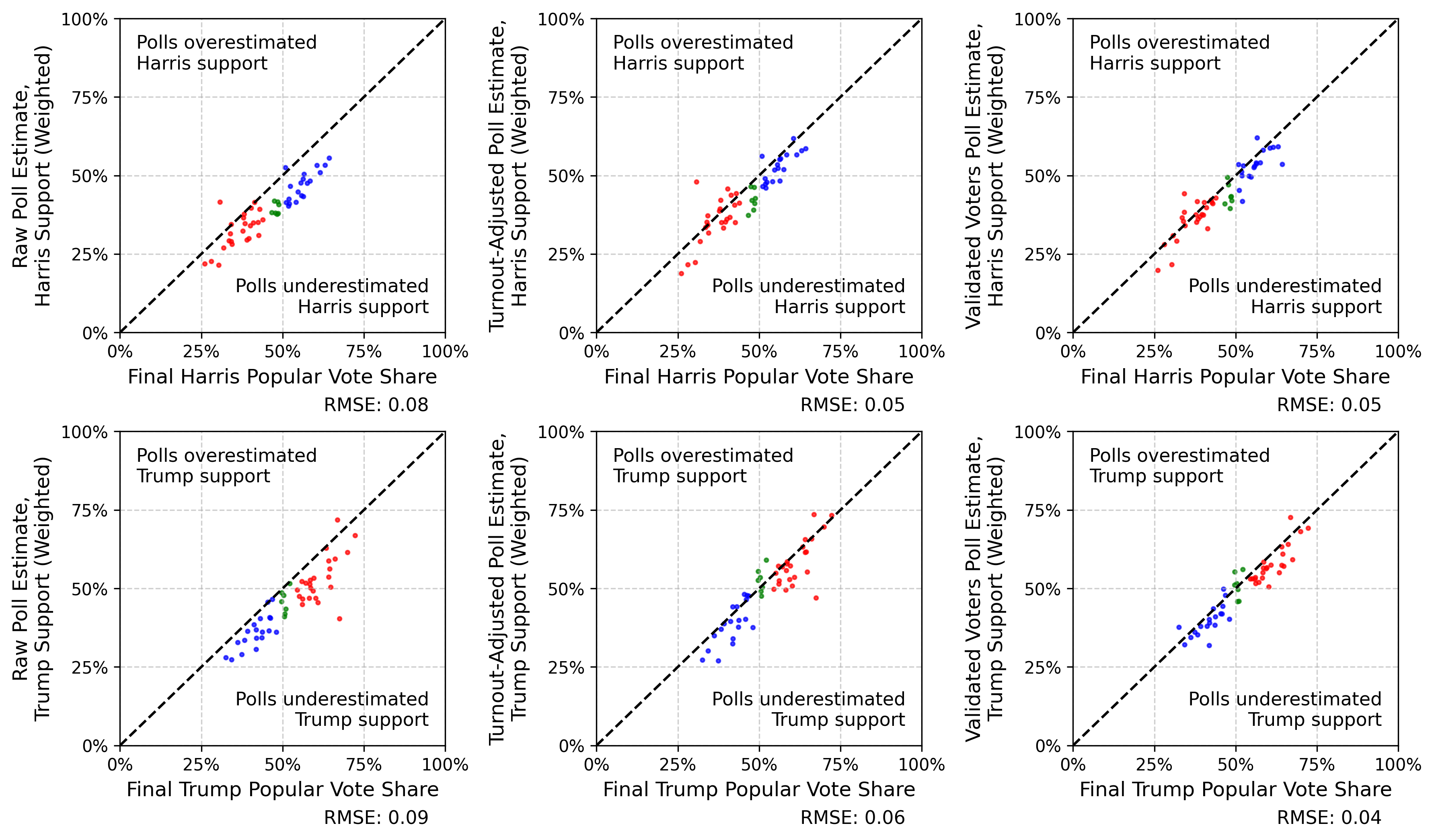}
  \caption{Polled vs true vote share of Trump and Harris using CES data, weighted. Note that the sum of the raw poll estimate does not sum up to 1 as 13.2\% of voters (across all states) chose one of ``Other'', ``Won't vote'', or ``Not sure''.}
  \label{fig:figure_4_weighted}
\end{figure}

In \Cref{fig:figure_4_weighted}, we plot the polled vs true vote share using weights computed by CES post-election. Weighting removes non-response bias, but requires assumptions including ignorability.

\section{Details on sample sizes and sample ratios}

\Cref{tab:sample-details} shows the sample ratio, sample size and total votes in each state for the 2024 US presidential election. Sample ratios and sample sizes are from CES \citep{schaffnerCooperativeElectionStudy2025}. The total votes are from \citet{federalelectioncommissionfecElectionResultsVoting2024}.

\begin{center}

\begin{longtable}{lrrr}
\caption{Sample ratio, sample size and total votes in each state for the 2024 US presidential election}\label{tab:sample-details}\\
\hline
State & Sample ratio & Sample size & Total votes \\
\hline
Alabama & 0.00039 & 882 & 2,256,352 \\
Alaska & 0.00035 & 117 & 338,177 \\
Arizona & 0.00034 & 1,162 & 3,389,319 \\
Arkansas & 0.00042 & 492 & 1,182,676 \\
California & 0.00025 & 4,035 & 15,862,536 \\
Colorado & 0.00027 & 867 & 3,190,873 \\
Connecticut & 0.00030 & 536 & 1,758,429 \\
Delaware & 0.00039 & 202 & 511,697 \\
Florida & 0.00033 & 3,543 & 10,893,548 \\
Georgia & 0.00031 & 1,608 & 5,250,047 \\
Hawaii & 0.00030 & 153 & 516,701 \\
Idaho & 0.00027 & 246 & 904,812 \\
Illinois & 0.00035 & 1,936 & 5,592,368 \\
Indiana & 0.00037 & 1,078 & 2,933,770 \\
Iowa & 0.00033 & 544 & 1,656,849 \\
Kansas & 0.00034 & 449 & 1,327,591 \\
Kentucky & 0.00041 & 844 & 2,073,309 \\
Louisiana & 0.00031 & 615 & 2,006,975 \\
Maine & 0.00027 & 224 & 831,375 \\
Maryland & 0.00029 & 868 & 3,015,650 \\
Massachusetts & 0.00027 & 917 & 3,453,369 \\
Michigan & 0.00026 & 1,470 & 5,662,504 \\
Minnesota & 0.00025 & 799 & 3,240,916 \\
Mississippi & 0.00039 & 475 & 1,228,008 \\
Missouri & 0.00041 & 1,215 & 2,993,596 \\
Montana & 0.00029 & 173 & 602,963 \\
Nebraska & 0.00030 & 285 & 947,159 \\
Nevada & 0.00040 & 590 & 1,484,840 \\
New Hampshire & 0.00034 & 283 & 822,116 \\
New Jersey & 0.00032 & 1,362 & 4,272,725 \\
New Mexico & 0.00035 & 321 & 923,403 \\
New York & 0.00039 & 3,205 & 8,199,062 \\
North Carolina & 0.00027 & 1,537 & 5,679,647 \\
North Dakota & 0.00027 & 99 & 365,059 \\
Ohio & 0.00036 & 2,083 & 5,765,017 \\
Oklahoma & 0.00043 & 672 & 1,566,173 \\
Oregon & 0.00040 & 891 & 2,229,467 \\
Pennsylvania & 0.00036 & 2,503 & 7,034,206 \\
Rhode Island & 0.00035 & 178 & 510,659 \\
South Carolina & 0.00035 & 895 & 2,548,140 \\
South Dakota & 0.00031 & 131 & 428,922 \\
Tennessee & 0.00037 & 1,133 & 3,063,942 \\
Texas & 0.00037 & 4,222 & 11,380,105 \\
Utah & 0.00029 & 426 & 1,487,951 \\
Vermont & 0.00023 & 85 & 366,389 \\
Virginia & 0.00026 & 1,177 & 4,482,576 \\
Washington & 0.00030 & 1,166 & 3,898,835 \\
West Virginia & 0.00053 & 403 & 762,390 \\
Wisconsin & 0.00032 & 1,077 & 3,415,213 \\
Wyoming & 0.00033 & 88 & 266,353 \\
\hline

\end{longtable}
\end{center}

\begin{figure}[!htb]
\centering
\includegraphics[width=.7\linewidth]{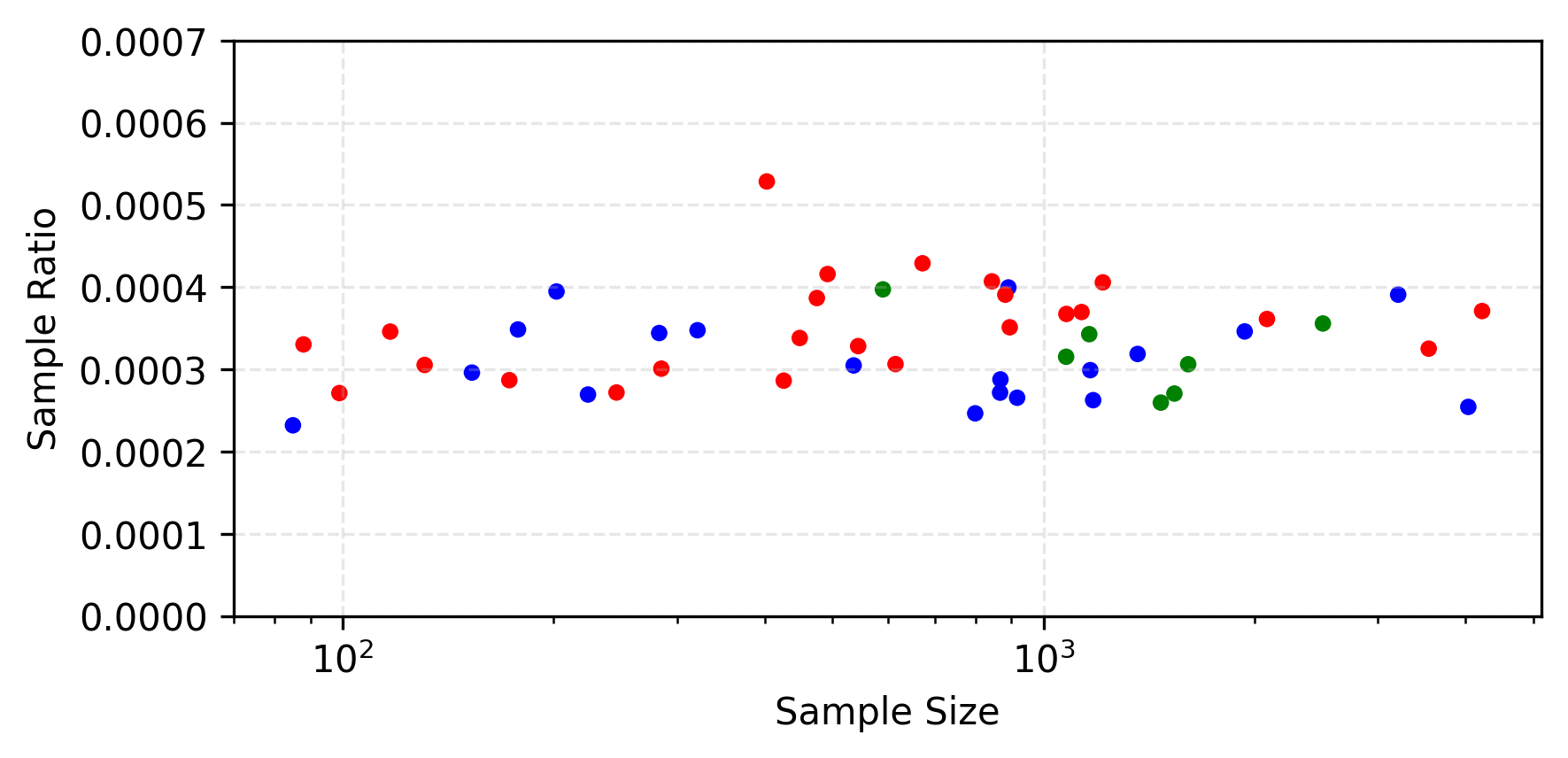}
\caption{Sample ratio to sample size of \cite{schaffnerCooperativeElectionStudy2025}. {The x-axis uses a log scale.} Note that the sample ratio stays relatively constant as sample size grows}
\label{fig:sample-size-ratio-scatterplot}

\end{figure}

\Cref{fig:sample-size-ratio-scatterplot} shows that sample ratios in CES do not vary as widely as much as the sample sizes for each state.

\section{Details on effective sample sizes}
\Cref{tab:effective-sample-sizes} shows the effective sample sizes and percentage reduction for CES \citep{schaffnerCooperativeElectionStudy2025} in each state for the 2024 US presidential election.

\begin{center}

\begin{longtable}{lrrrr}
\caption{Sample size, total votes, effective sample size and percentage reduction of samples in each state for the 2024 US presidential election}\label{tab:effective-sample-sizes}\\
\hline
State & Sample size & Total votes & Effective sample size & Percentage reduction \\
\hline
Alabama & 882 & 2,256,352 & 20 & 97.71\% \\
Alaska & 117 & 338,177 & 18 & 84.72\% \\
Arizona & 1,162 & 3,389,319 & 18 & 98.48\% \\
Arkansas & 492 & 1,182,676 & 21 & 95.63\% \\
California & 4,035 & 15,862,536 & 13 & 99.67\% \\
Colorado & 867 & 3,190,873 & 14 & 98.38\% \\
Connecticut & 536 & 1,758,429 & 16 & 97.06\% \\
Delaware & 202 & 511,697 & 20 & 89.90\% \\
Florida & 3,543 & 10,893,548 & 17 & 99.53\% \\
Georgia & 1,608 & 5,250,047 & 16 & 99.02\% \\
Hawaii & 153 & 516,701 & 15 & 90.00\% \\
Idaho & 246 & 904,812 & 14 & 94.29\% \\
Illinois & 1,936 & 5,592,368 & 18 & 99.08\% \\
Indiana & 1,078 & 2,933,770 & 19 & 98.24\% \\
Iowa & 544 & 1,656,849 & 17 & 96.88\% \\
Kansas & 449 & 1,327,591 & 17 & 96.11\% \\
Kentucky & 844 & 2,073,309 & 21 & 97.51\% \\
Louisiana & 615 & 2,006,975 & 16 & 97.43\% \\
Maine & 224 & 831,375 & 14 & 93.79\% \\
Maryland & 868 & 3,015,650 & 15 & 98.29\% \\
Massachusetts & 917 & 3,453,369 & 14 & 98.50\% \\
Michigan & 1,470 & 5,662,504 & 13 & 99.09\% \\
Minnesota & 799 & 3,240,916 & 13 & 98.41\% \\
Mississippi & 475 & 1,228,008 & 20 & 95.79\% \\
Missouri & 1,215 & 2,993,596 & 21 & 98.27\% \\
Montana & 173 & 602,963 & 15 & 91.43\% \\
Nebraska & 285 & 947,159 & 16 & 94.54\% \\
Nevada & 590 & 1,484,840 & 21 & 96.52\% \\
New Hampshire & 283 & 822,116 & 18 & 93.71\% \\
New Jersey & 1,362 & 4,272,725 & 16 & 98.79\% \\
New Mexico & 321 & 923,403 & 18 & 94.40\% \\
New York & 3,205 & 8,199,062 & 20 & 99.37\% \\
North Carolina & 1,537 & 5,679,647 & 14 & 99.09\% \\
North Dakota & 99 & 365,059 & 14 & 85.85\% \\
Ohio & 2,083 & 5,765,017 & 19 & 99.10\% \\
Oklahoma & 672 & 1,566,173 & 22 & 96.70\% \\
Oregon & 891 & 2,229,467 & 21 & 97.68\% \\
Pennsylvania & 2,503 & 7,034,206 & 18 & 99.27\% \\
Rhode Island & 178 & 510,659 & 18 & 89.88\% \\
South Carolina & 895 & 2,548,140 & 18 & 97.97\% \\
South Dakota & 131 & 428,922 & 16 & 87.95\% \\
Tennessee & 1,133 & 3,063,942 & 19 & 98.31\% \\
Texas & 4,222 & 11,380,105 & 19 & 99.55\% \\
Utah & 426 & 1,487,951 & 15 & 96.53\% \\
Vermont & 85 & 366,389 & 12 & 85.90\% \\
Virginia & 1,177 & 4,482,576 & 14 & 98.85\% \\
Washington & 1,166 & 3,898,835 & 15 & 98.67\% \\
West Virginia & 403 & 762,390 & 27 & 93.22\% \\
Wisconsin & 1,077 & 3,415,213 & 16 & 98.49\% \\
Wyoming & 88 & 266,353 & 17 & 80.60\% \\
\hline

\end{longtable}
\end{center}

\section{Plotting \(Z_n\) with all respondents and likely voters}\label{sec:figures-using-all}

We replot the standardized error figure using all respondents and likely voters. The same pattern holds where \(Z_n\) tends to escape the confidence interval for larger samples.

\begin{figure}[!htb]
\centering
\includegraphics[width=\linewidth]{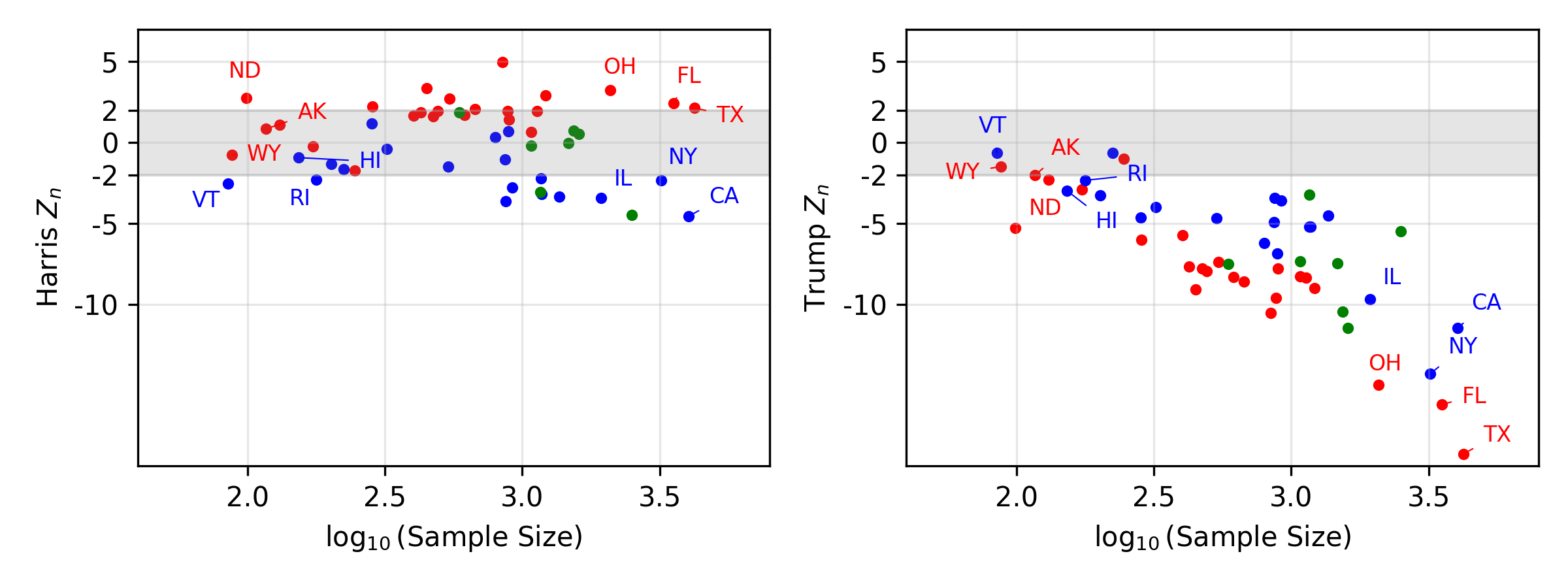}
\caption{Standardized error \(Z_n\) to sample size for each state, using all respondents.}
\label{fig:all-fig-7-big-data-paradox-all}
\end{figure}

\begin{figure}[!htb]
\centering
\includegraphics[width=\linewidth]{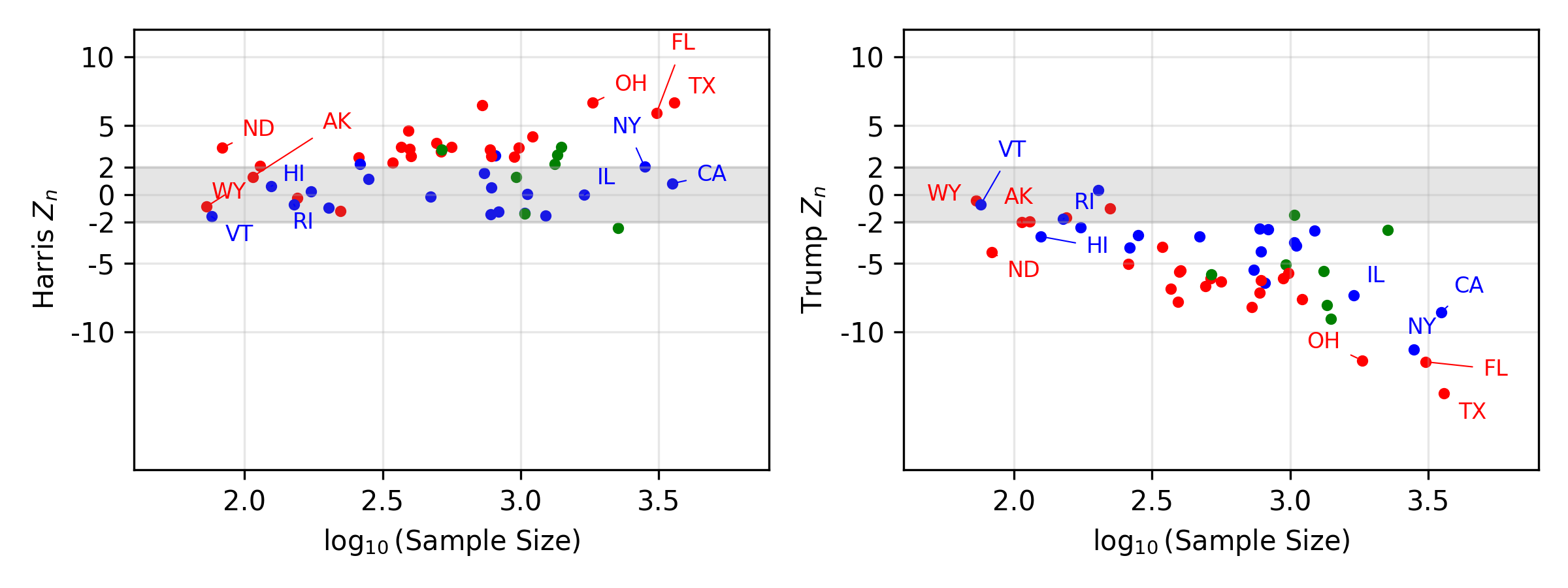}
\caption{Standardized error \(Z_n\) to sample size for each state, using likely voters.}
\label{fig:all-fig-7-big-data-paradox-likely}
\end{figure}

\section{Accounting for dishonest responses}
\begin{figure}[!htb]
\centering
\includegraphics[width=\linewidth]{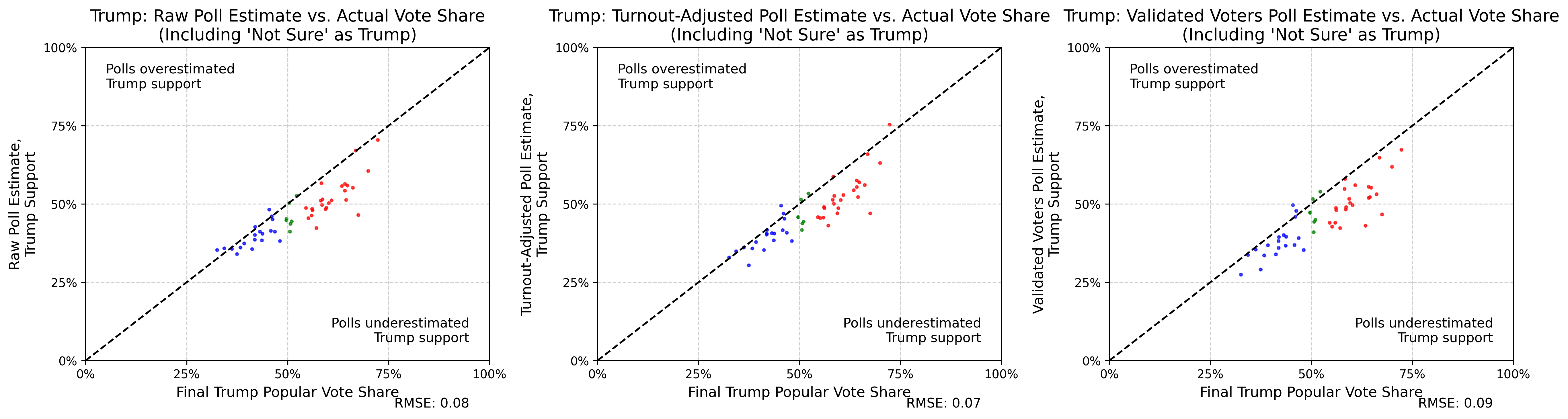}
\caption{Poll estimate vs actual vote share for Trump for each state, including ``I'm not sure'' as part of the Trump estimate}
\label{fig:not-sure-is-trump}
\end{figure}

In \Cref{fig:not-sure-is-trump}, we replotted the poll estimate vs actual vote share for Trump but delegating ``I'm not sure'' responses as part of Trump's estimate. This decreased the RMSE and gave a better fit, but did not match the performance of the polls for Harris or that of the bias-corrected estimator.

\section{Details on estimating turnout}\label{subsec:turnout-estimate-details}
To estimate turnout, we coarsen the data on the voting intention data in CES dataset and consider those who expressed any voting intention (i.e. one of ``Yes'', ``Probably'', ``Voted'') as likely voters, and discard the responses that responded ``No'' or ``Undecided''.
\begin{equation}
  \bar{G}_{n,\text{turnout-adjusted}}
  = \frac{\text{Number of likely voters who supports the candidate}}{\text{Total number of likely voters}}
\end{equation}
A more sophisticated approach will be to apply weighting to each of ``Yes'', ``Probably'', ``Voted'', but it is unclear how much weight should be assigned to each class of voters. Resampling from those who chose ``Probably'' or ``Undecided'' has been considered, but again there is no clear principle on what probabilities to assign to these classes.

\end{document}